# Theoretical Exploration of the Diene-Transmissive Hetero-Diels-Alder Strategy Toward Boron-Functionalized Octahydroquinolines

Amine Rafik[a,b], Abdeljabbar Jaddi[a], Kahlid Abbiche[a], Mohammed Salah[c] Miguel Carvajal[b], and Khadija Marakchi*[a]

a Laboratory of Spectroscopy, Molecular Modeling, Materials, Nanomaterials, Water and Environment, LS3MN2E/CERNE2D, Faculty of Sciences, Mohammed V University in Rabat, Morocco, k.marakchi@um5r.ac.ma.
b Departamento de Ciencias Integradas, Centro de Estudios Avanzados en Física, Matemática y Computación; Unidad Asociada GIFMAN, CSIC-UHU, Universidad de Huelva, Huelva, 21071, Spain.
c Molecular Modelling and Spectroscopy Research Team, Faculty of Science, Chouaïb Doukkali University, P.O. Box 20, 24000 El Jadida, Morocco.

## Abstract

A diene-transmissive hetero-Diels–Alder strategy, grounded in previous experimental works and employing boronated dienophiles, is proposed for the synthesis of boron-bearing octahydroquinolines. To assess its feasibility, three representative reactions were investigated, and their thermodynamics were evaluated in toluene and acetonitrile at various temperatures using the ωB97X-D level of theory. The peri-, regio-, stereo-, and π-facial selectivities were predicted. The reactions mechanisms were elucidated through exploration of the reaction pathways. The predictions are consistent with available experimental work, and show the reactions are feasible with low to moderate polarity. The results also demonstrate that the reactions selectivity can in some cases be tuned by judicious choice of reaction conditions to deliver specific products with high selectivity.

**Keywords**: Octahydroquinolines, Organoboron compounds, Diels-Alder reaction, DFT, MEDT.

# Introduction

Quinolines occupy a privileged position in organic and medicinal chemistry due to their versatile physicochemical properties and broad spectrum of biological activities [1, 2]. Their diversity underscores how subtle modifications to their structure can significantly their properties, particularly when guided with modern computational methods [3, 4]. By contrast, their hydrogenated analogs, polyhydroquinolines, are underexplored, usually due to their lack of a conjugated planar structure supposedly important for various biological interactions. They also have lower synthetic accessibility, especially their highly hydrogenated analogues (i.e., octahydroquinolines). They are nevertheless interesting scaffolds that deserve more attention [5–9]. Moreover, polyhydroquinolines have a rich chiral chemistry, and they fall well within the "escape-from-flatland" framework which stipulates that saturation correlates well with important pharmacological properties, especially solubility [10].

Organoboron compounds have emerged as both indispensable synthetic intermediates and promising bioactive motifs. Boronic acids and boronate esters uniquely combine Lewis acidity with reversible covalent binding, a feature exploited in cross-coupling reactions (e.g., Suzuki-Miyaura) and in pharmaceutical agents [11–13]. Building on these patterns, we propose the design of octahydroquinoline derivatives that bear a boron moiety. Furthermore, the saturated quinoline core incorporates a sulfonyl group, well known for its desirable effects in medicinal chemistry [14]. Combining saturated quinoline frameworks with unique functional motifs, such as boron groups or sulfonyl substituents, could yield scaffolds with enhanced properties beyond those of known bioisosteres of quinoline. Such frameworks remain a largely untapped chemical space. The convergent assembly of such substituted, partially saturated quinoline cores presents two principal challenges. First, access to polyhydroquinoline frameworks often requires multistep hydrogenation under precise conditions to ensure regio- and stereochemical control, or alternatively the recently developed late-stage saturation methodology [15]. Second, the site-selective installation of boronic acids or boronate esters on a dense, sp³-rich skeleton is hindered by steric congestion and competing C-H activation pathways; existing methods (e.g., Ir- or Pd-catalyzed C-H borylation under high temperatures or strong Lewis acids) perform variably on heterocyclic substrates and sometimes struggle to deliver enantioselective outcomes [16–18].

To surmount these obstacles, we adopt a two-step cascade approach based on the Diene-Transmissive Hetero-Diels-Alder (DTHDA) reaction [19–23]. In this approach, an activated heterodiene, typically a dendralene (e.g., an azatriene), first undergoes a cycloaddition with a dienophile, forming an intermediate that transmutes into a second diene for subsequent cycloaddition which eventually leads to a fused compound as shown in **Scheme 1**. This approach is highly efficient considering that the Diels-Alder (DA) reaction is conceptually one of the simplest organic reactions possible, although the mechanism is not as simple as it was traditionally believed to be [24–26]. The approach has been applied employing various dienes and dienophiles [19, 20, 23, 27] and an analogous approach was recently designed for 1,3-dipolar cycloadditions [28]. Recently, the use of organoboron compounds in pericyclic reactions has been reviewed [29]. A wide range of dienes and dienophiles can be employed, allowing for the introduction of boron-containing groups as part of the diene or the dienophile. One commonly reported trend is the necessity of a Lewis-acid catalyst. In any case, due to the highly chemo-, regio-, and stereoselective nature of the DA reaction, it offers a promising alternative to traditional synthetic methods of boron-bearing quinolines, addressing many of their inherent challenges.

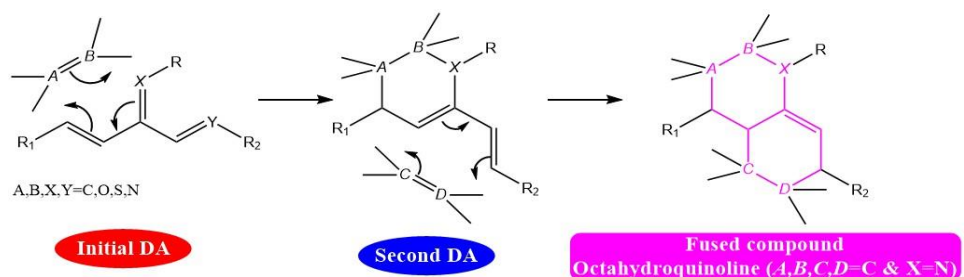

*Scheme 1: General scheme of a Diene-Transmissive-Diels-Alder reaction*

In the context of the current employed DTHDA approach, we employ three structurally distinct and synthetically accessible boron-substituted dienophiles (**Scheme 2**), namely, allenylboronic acid pinacol ester (**a**) [30], the alkylhalovinylboranes, chlorocyclohexylvinylborane (**b1**) and bromocyclohexylvinylborane (**b2**) [31], and 4S,5S-[Bis(carbethoxy)]-2-ethenyl-1,3,2-dioxaborolane (**c**) [32]. Dienophiles **a** and **c** contain a dioxaborolane functional group which has been highlighted in the context of drug design [33–35]. In this work, we explore the possibility of incorporating these dienes either in the first or the second step of the DTHDA reaction.

In a recent work, we have elucidated the mechanistic details related to the aza-DA reaction of substituted derivatives of an azatriene with the dienophiles ethyl vinyl ether and allenyl methyl ether. The computational insights have shown good agreement with the experimental observations [36]. Here we consider two possible scenarios, in the first (Route 1 in **Scheme 2**), the B-containing dienophiles react with an azatriene diene in the first step of the DTHDA procedure and then they react again with the product from the first step, in which case we will obtain boron-rich octahydroquinoline products where the boron functional groups have been incorporated both in the pyridine and the cyclohexene rings. Previous works suggest that this route is little accessible [30, 36, 37] but for a systematic investigation of the reactivity of the system we will explore this option by characterizing the reactants using conceptual Density Functional Theory and proceed according to the results. In the second scenario (Route 2 in **Scheme 2**), the dienophiles react in the second step of the DTHDA procedure, in which case, different dienophiles ought to be used in the first step and it is already established that ethyl vinyl ether (alternatively ethyl vinyl sulfide or allenyl methyl ether) is an experimentally validated dienophile for this task [21]. **Scheme 2** illustrates the two possible scenarios. We note that the products from the reaction of azatriene with ethyl vinyl ether are generally labeled **VPS** (owing to the Pyridine core, and the Vinyl and Sulfonyl substituents shared among all products) followed by an index, while the products from the reaction of azatriene with the B-containing dienophiles are generally labeled as **BVPS**. **VPS** (without an index) also represents the main diene (**Scheme 2**), but upon substitution of R in the $SO_2$ group (R=Me, Ph, p-Tol, PhCO), the X group (X=OEt, SEt), and the Y group (Y=$H_2$, $CH_2$), we obtain eleven more diene derivatives (labeled **VPS2-12**) that are also considered in the analysis. The labeling of the products of the second step in the DTHDA procedure is provided in the following sections. Initially, we explore the feasibility of the first step of the DTHDA approach using the azatriene diene and the proposed dienophiles. Following that we proceed to the second step of the DTHDA approach. The reactions myriad can yield multiple isomeric octahydroquinolines, distinguished by substituents, regio-, stereo-, and π-facial chemistry. The possible reaction pathways are depicted in the following sections.

To guide and predict the outcomes of these cascade processes, we employ an in silico workflow grounded in exploring reaction energy pathways, Conceptual Density Functional Theory (CDFT) and Molecular Electron Density Theory (MEDT). Such approach has been heavily benchmarked against experimental reactions and has shown excellent results [36, 38–47].

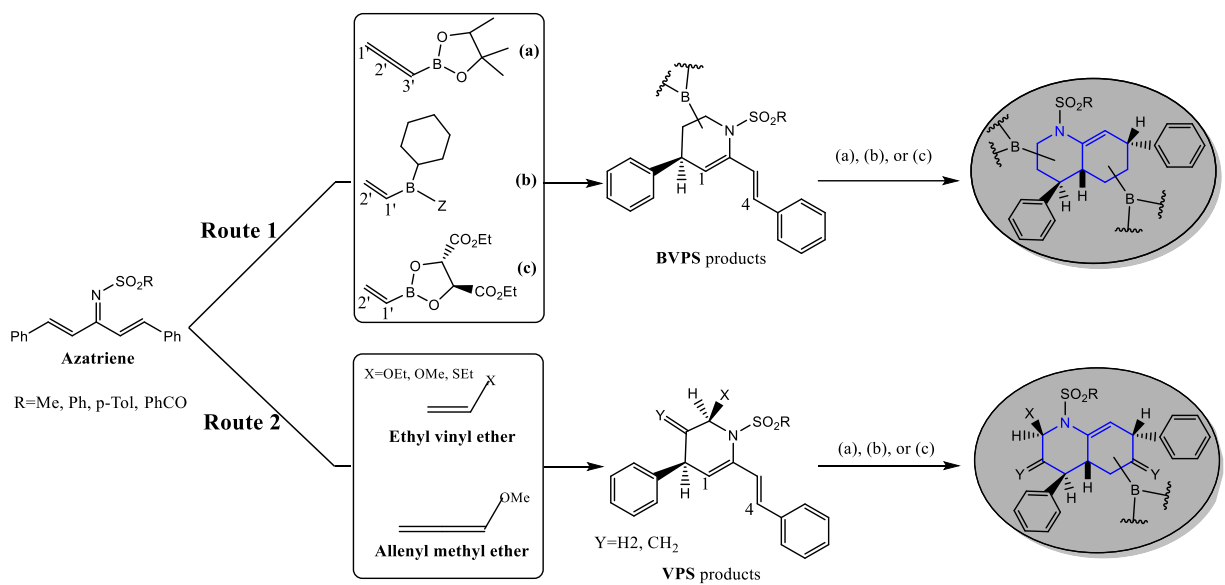

*Scheme 2: General scheme of the proposed reaction with the possible products*

## Computational Details

The ωB97X-D/6-311G(d,p) level of theory, as implemented in the Gaussian 09 package, is employed in all computations in the present work. According to recent studies, this functional is reliable for reactions energies [48], and generally in optimization jobs [49, 50]. Additionally, benchmarking it against a similar Diels-Alder work has further confirmed its reliability [36], and it has shown good results in the study of cycloadditions [48, 51, 52]. Initial reactants were subjected to a conformational search around relevant internal coordinates to determine the most stable conformers, and these were used for subsequent calculations. The reaction pathways were mapped by Intrinsic Reaction Coordinate calculations. The self-coherent reaction field (SCRF) is used to implicitly simulate the effect of the solvent at temperatures 25 °C, 70 °C, and 100 °C. The workflow employed here has been meticulously reported in other studies [36, 47, 53, 54], in here we stick to a minimal description. CDFT and the Global Electron Density Transfer (GEDT) analyses has been applied as described by Domingo et al [55–57]. Electron density analysis was carried out using the dependent Gradient Model based on Hirshfeld partition of molecular density (IGMH) variant [58] of the IGM [59], and the Electron Localization Function (ELF) method [60]. Multiwfn 3.8 was used for the IGMH calculations [61] and TopMod was used for the ELF ones [62]. Visualization software, GaussView and Visual Molecular dynamics (VMD) were used to prepare the various figures [63, 64].

## Results and Discussion

### 1. Characterization of the reactants

#### 1.1. Global electrophilicity and nucleophilicity analysis

CDFT is a useful approach to the absolute reactivity of chemical reactants. The electrophilicity (ω) and nucleophilicity (N) scales derived within CDFT are grounded in a theoretical framework that captures the fundamental electronic characteristics of molecules and maintains good correlation with experimental observations of chemical reactivity [65, 66]. In the context of the employed level of theory, on the one hand, molecules are termed weak, moderate, or strong nucleophiles if their calculated N is below 1.91 eV, between 1.91 eV and 2.93 eV, or higher than 2.93 eV. On the other hand, they are labeled as weak, moderate, or strong electrophiles if their calculated ω index is below, 0.59 eV, between 0.59 eV and 0.98 eV, or higher than 0.98 eV, respectively. The results of the CDFT indices on the isolated reactants are summarized in **Table 1**. We recall that per experimental data, azatriene is an electron-deficient diene that primarily reacts with electron-rich dienophiles like ethyl vinyl ether [20]. According to the calculated global indices, azatriene is actually both a strong electrophile and strong nucleophile while ethyl vinyl ether is weak electrophile and strong nucleophile. In contrast, none of the boron-bearing dienophiles **a**, **b1**, **b2**, and **c** shows a strong nucleophilic character typical of electron-rich dienophiles. In fact, except for **a** which is a moderate nucleophile, the rest are weak nucleophiles. Therefore, considering the first step of the proposed cascade reaction, we expect that the boronated dienophiles are not compatible with azatriene unless under harsh reaction conditions, and consequently we omit Route 1 in **Scheme 2**. For Route 2 in **Scheme 2**, the experimental and theoretical results for the first step 1 of the DTHDA reaction are well known and the reader can consult them in references [21, 36]. However, for comparison purposes, we include the global CDFT results for ethyl vinyl ether and allenyl methyl ether in **Table 1**.

Considering the reaction of the B-containing dienophiles with **VPS** in the second step of the DTHDA reaction, we see that the dienophiles all belong to a different class in comparison to **VPS** at least in one of the two scales of electrophilicity or nucleophilicity. This indicates that the reactants have favorable electrophilic/nucleophilic interactions, and consequently, Route 2 in **Scheme 2** is accessible. In particular, **VPS** is a strong nucleophile while the boron-bearing dienophiles, on the one hand, show weak to moderate nucleophilicity and, on the other hand, moderate (**a** and **c**) to strong (**b1** and **b2**) electrophilicity. In the experimental work of Labadie et al. [30] we see that **a** reacts readily in a Diels-Alder reaction with cylopentadiene. Inspection of their CDFT indices reveals that the cyclopentadiene is a strong nucleophile with N=3.27 eV. Hence, it is classified similar the diene employed in this work and we expect that the reaction considered here is promising. Besides the main diene (**VPS**), all other variants (**VPS2-12**) show similar values to the main diene, confirming that from an absolute reactivity point of view, the reactants will interact favorably to form new cycloadducts, especially in the case of dienophiles **b1** and **b2** with **VPS**, and any possible low reactivity can be improved by the right solvent and temperature.

*Table 1: Global CDFT indices for all dienes and dienophiles with all values given in units of eV.*

|  | HOMO | LUMO | μ | η | ω | N |
|---|---|---|---|---|---|---|
| **azatriene** | -8.44 | -1.05 | -4.74 | 7.39 | 1.52 | 2.96 |
| **VPS** (X=O, Y=H2, R=Me) | -7.86 | 0.26 | -3.80 | 8.13 | 0.89 | 3.53 |
| **VPS2** (X=O, Y=H2, R=PhSO2) | -7.9 | 0.21 | -3.86 | 8.15 | 0.91 | 3.46 |
| **VPS3** (X=O, Y=H2, R=TolSO2) | -7.9 | 0.24 | -3.84 | 8.15 | 0.91 | 3.48 |
| **VPS4** (X=O, Y=H2, R=PhCO) | -7.7 | 0.34 | -3.66 | 8 | 0.84 | 3.74 |
| **VPS5** (X=S, Y=H2, R=Me) | -7.98 | 0.22 | -3.88 | 8.2 | 0.92 | 3.41 |
| **VPS6** (X=S, Y=H2, R=PhSO2) | -7.95 | 0.14 | -3.91 | 8.1 | 0.94 | 3.44 |
| **VPS7** (X=S, Y=H2, R=TolSO2) | -8 | 0.15 | -3.93 | 8.15 | 0.95 | 3.4 |
| **VPS8** (X=S, Y=H2, R=PhCO) | -7.7 | 0.25 | -3.73 | 7.96 | 0.87 | 3.69 |
| **VPS9** (X=O, Y=CH2, R=Me) | -7.69 | 0.42 | -3.64 | 8.11 | 0.82 | 3.71 |
| **VPS10** (X=O, Y=CH2, R=PhSO2) | -7.78 | 0.36 | -3.71 | 8.14 | 0.85 | 3.61 |
| **VPS11** (X=O, Y=CH2, R=TolSO2) | -7.75 | 0.38 | -3.69 | 8.13 | 0.84 | 3.65 |
| **VPS12** (X=O, Y=CH2, R=PhCO) | -7.71 | 0.14 | -3.79 | 7.85 | 0.91 | 3.69 |
| **Ethyl vinyl ether** | -8.25 | 2.76 | -2.74 | 11.01 | 0.34 | 3.15 |
| **Allenyl methyl ether** | -8.29 | 1.83 | -3.23 | 10.1 | 0.51 | 3.11 |
| **a** | -9.24 | 1.48 | -3.88 | 10.7 | 0.70 | 2.15 |
| **b1** | -9.72 | 0.04 | -4.84 | 9.75 | 1.20 | 1.68 |
| **b2** | -9.59 | -0.03 | -4.81 | 9.57 | 1.21 | 1.80 |
| c | | -9.61 | 1.24 | -4.18 | 10.8 | 0.81 | 1.79 |

## 1.2. Local electrophilicity and nucleophilicity analysis

Periselectivity and regioselectivity are two important aspects of chemical synthesis. All considered reactions in this work present a regioselectivity aspect, and the species **a**, being an allene, also has a periselectivity aspect. Parr functions, $P^+_k$ and $P^-_k$, allow us to evaluate the electrophilicity and nucleophilicity at the local level of the atoms of the reacting species. These local indices make it possible to predict the preferred bonding sites from each reactant and, therefore, peri- and regioselectivity when applicable. These local indices have been calculated for the diene and dienophiles considered in this work and are reported in **Table 2**. We note that since all substituted dienes have shown similar values, in here we only report the main one, i.e., **VPS**.

*Table 2: Parr functions and local CDFT indices for VPS and all dienophiles. All values given in units of eV.*

| | Atomic site | $P^+_k$ | $P^-_k$ | $\omega_k$ | $N_k$ |
|---|---|---|---|---|---|
| 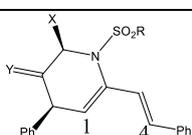 **VPS** | C1 | 0.314688 | 0.422844 | 0.28 | 1.49 |
| | C2 | 0.194313 | 0.176116 | 0.17 | 0.62 |
| 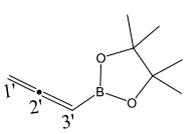 **a** | C1 | -0.114568 | 0.247882 | -0.08 | 0.53 |
| | C2 | 0.518138 | 0.190590 | 0.37 | 0.41 |
| | C3 | 0.093426 | 0.293847 | 0.07 | 0.63 |
| 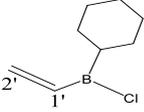 **b1** | C1 | -0.094718 | 0.067735 | -0.11 | 0.11 |
| | C2 | 0.503374 | 0.035035 | 0.61 | 0.06 |
| 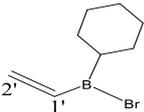 **b2** | C1 | -0.108746 | 0.066766 | -0.13 | 0.12 |
| | C2 | 0.487270 | 0.008735 | 0.59 | 0.02 |
| 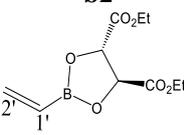 **c** | C1 | 0.072713 | 0.123424 | 0.06 | 0.22 |
| | C2 | 0.528320 | 0.133959 | 0.43 | 0.24 |

The results indicate that the most nucleophilic center in **VPS** is the C1 carbon with a local nucleophilicity index of 1.49. In the context of the reaction of **VPS** and **a**, it is predicted that the favored bonding will involve the proximal double bond of **a**, where C1 ($N_k$=1.49) bonds with C2' ($\omega_k$=0.37) and C2 ($N_k$ = 0.62) bonds with C3' ($\omega_k$ = 0.07). It is worthy of mentioning that in the works of Labadie et al 2021 and Labadie et al 2022 [30, 37], which involves the same dienophile and an electronically similar diene to **VPS**, their CDFT analysis also indicates that bonding at the proximal double bond is favored, and their experimental work confirms it. In the reactions of **VPS** with **b1** and **b2**, both the chlorine and the bromine substituted dienophiles exhibit similar regioselectivity. In both cases, the main interaction is between atoms C1 from **VPS** and C2' from the dienophiles ($\omega_k$ =

0.61/0.59 for Cl and Br), where C2' is the boron-substituted carbon. Products arising from this regioselectivity will be labeled with an *r2* label.

The dienophile from reaction 3 also shows similar local indices and, therefore, a similar bonding pattern is proposed where the favorable bonding interaction is C1 of the diene with C2' ($\omega_k = 0.43$) of the dienophile. We use *r2* to label this regioselectivity in the stationary points.

### 1.3. Electron Localization Function analysis of the reactants

The ELF provides a quantum-mechanical framework that translates the familiar Lewis model of localized electron pairs into an observable picture making it possible to trace bond breaking, bond formation, and charge transfer with great detail. The ELF is a real-space indicator of how "localized" electrons are at each point in a molecule. Maxima of the ELF define attractors and the gradient of ELF partitions space into basins surrounding each attractor. Each basin corresponds chemically to a lone pair, covalent bond, core shell, or more delocalized interaction. Integrating over these basins can provide electron populations which, in our case, allow us to analyze the electronic characteristics of the reactants. The calculated basin attractor positions, and the selected valence basin populations are shown in **Figure 1**.

The diene reveals populations consistent with the expected Lewis bonding framework as well as highlights the real-space polarization of electron density. The C1=C2 and C3=C4 double bonds show disynaptic basins with populations that sum to approximately 3.6 e and 3.4 e, respectively, slightly below the idealized 4 e. This indicates partial π-electron redistribution leading to small electronic impoverishment of the C=C double bonds in the 4 π-electron nucleophilic part of the diene. The sulfonyl group is a characteristic part of the diene that it is expected to take part in various interactions. Therefore, it is interesting to inspect it. It is revealed the S-O double bonds show a characteristic partitioning: each S-O disynaptic basin holds about 1.9 e, corresponding to the σ-bond, while the oxygen centers host two monosynaptic basins of about 2.7-3.2 e each. This pattern reflects the strong polarization of the π-electron density toward the oxygen atom. This is common in polar multiple bonds like S=O or C=O, where the π-electrons are not evenly shared but instead become part of the lone pair environment around the more electronegative atom.

The basin populations for the dienophile **a** reveal clear signatures of electron delocalization. For the two adjacent double bonds of the dienophile **a**, the analysis shows that the distal double features a disynaptic basin V(C1,C2) with a population of 3.72 e, while the proximal double bond shows a disynaptic basin V(C2,C3) summing to 3.50 e. This depletion, which is marginally stronger in the case of the proximal double bond, indicates that the π-electron density is partially lost and, consequently, the C2=C3 double bond is activated toward a nucleophilic attack by the diene, supporting the earlier conclusion from the Parr indices. This finding indicates that the major product from this reaction will arise from a DA reaction in the proximal bond. The electron density depletion in the C2=C3 bond can be traced back to the electron-withdrawing effect of the adjacent boron atom, whose empty p-orbital attracts π-electron density when the boron is tricoordinate. This is manifested by a disynaptic basin, V(C3,B), with a population of 2.34 e.

For the dienophiles **b1** and **b2**, very similar results are obtained for both and, therefore, we show only the case of the chlorine substituted dienophile. The dienophile **b1** shows disynaptic basins V(C1,C2) (1.61 e) and V(C1,C2) (1.58 e) with a total population of 3.19 e well below what is expected for an ideal double bond. It confirms the electrophilic nature predicted earlier by the electrophilicity index ω. This is due to the electron attractive effect of the adjacent boron as well as the highly electronegative chlorine center. In particular, the disynaptic basin V(C1,B) contains a population of 2.36 e and V(B,Cl) has a population of 1.68 e. These facts highlight the flow of electronic density from the double bond region to the adjacent environment.

Dienophile **c** shows two V(C1,C2) disynaptic basins of 1.60 e and 1.62 each, summing to 3.22 e. Furthermore, the C2-B bond is characterized by a population of 2.35 while the B-O bond in the dioxaborolane ring has a population of roughly 2 e. This shows that the dioxaborolane group works as an electron withdrawing group that drives the electrophilic behavior of **c**.

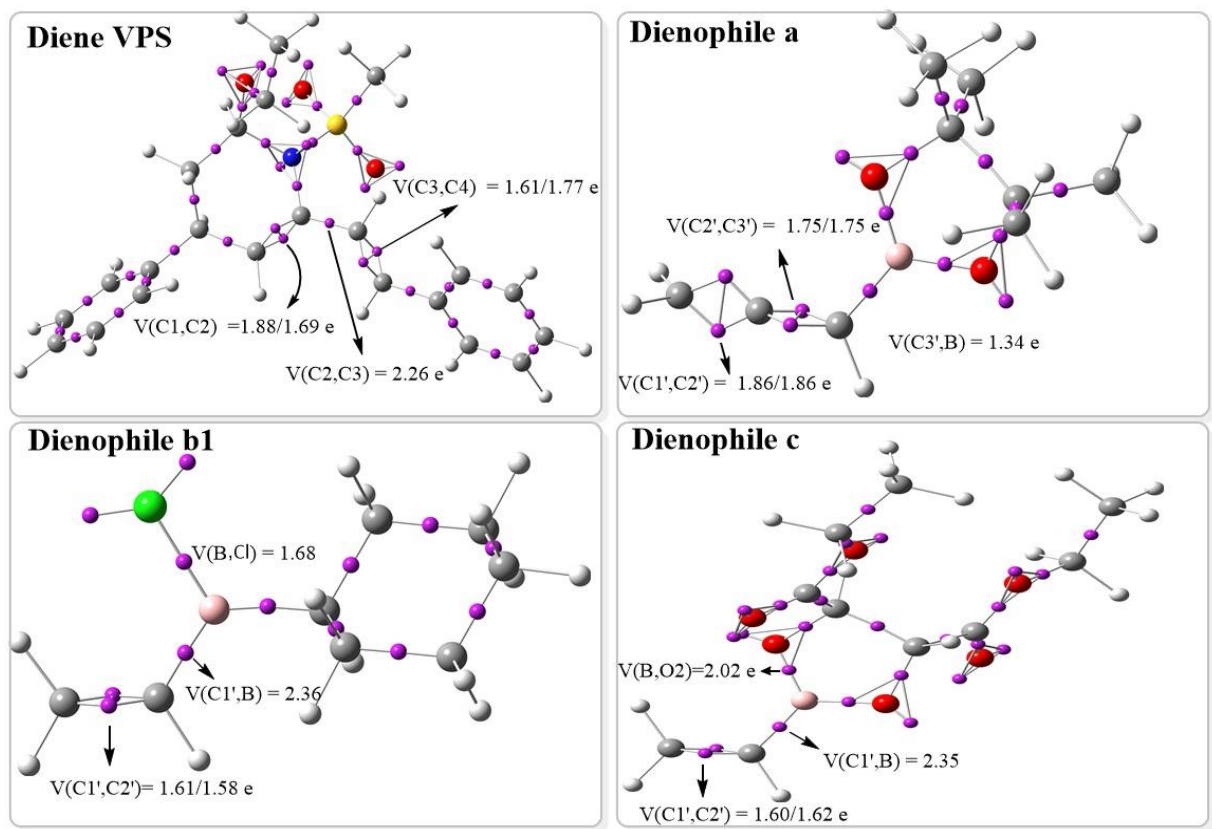

*Figure 1: ELF basin attractors for the reactants **VPS**, **a**, **b1**, and **c**.*

## 2. Energy analysis

We explore the stability of the possible reaction pathways for each reaction in the gas phase, and in the toluene and acetonitrile solvents. When in solvent, we also evaluate the effect of temperature at 70 °C and 100 °C in addition to the room temperature reference. In the following subsections we report the results and analysis for each reaction.

### 2.1. Energy analysis in the reaction of VPS with dienophile a

The reaction of **VPS** with **a** involves periselectivity. It can proceed according to two main routes relative to a [4+2] addition either on the proximal or distal double bond because the dienophile **a** belongs to the allenes family. Each of these two routes can involve regio-, stereo- and π-facial-isomers, leading to a total of sixteen possible reaction pathways. In the following sections, pathways relative to addition at the proximal double bond are labeled *prox* while those at the distal double bond are labeled *dis*. Additionally, we adopt a notation in which we specify the regio-, stereo- (or geometric configuration when applicable) and π-facial chemistry. The labels *r1* and *r2* refer to regioisomers associated with bonding at C1-C3' and C1-C2', respectively. The labels *ex*/*en* are used for the exo and endo approaches in the case of the *prox* approach while E/Z are used for the geometric configuration in the case of the *dis* approach. Finally, *f1* and *f2* are used to label the diastereo-π-facial selectivity. In particular, *f1* refers to an approach from under the **VPS** diene (as represented throughout the paper) and *f2* refers to an approach from above it. All possible products relative to the *f1* π-facial approach are illustrated in **Figure 2**. The possible products obtained from the *f2* approach are not shown but can be obtained by simply switching the stereochemistry at the asymmetric carbons C1 and C4 of **VPS**.

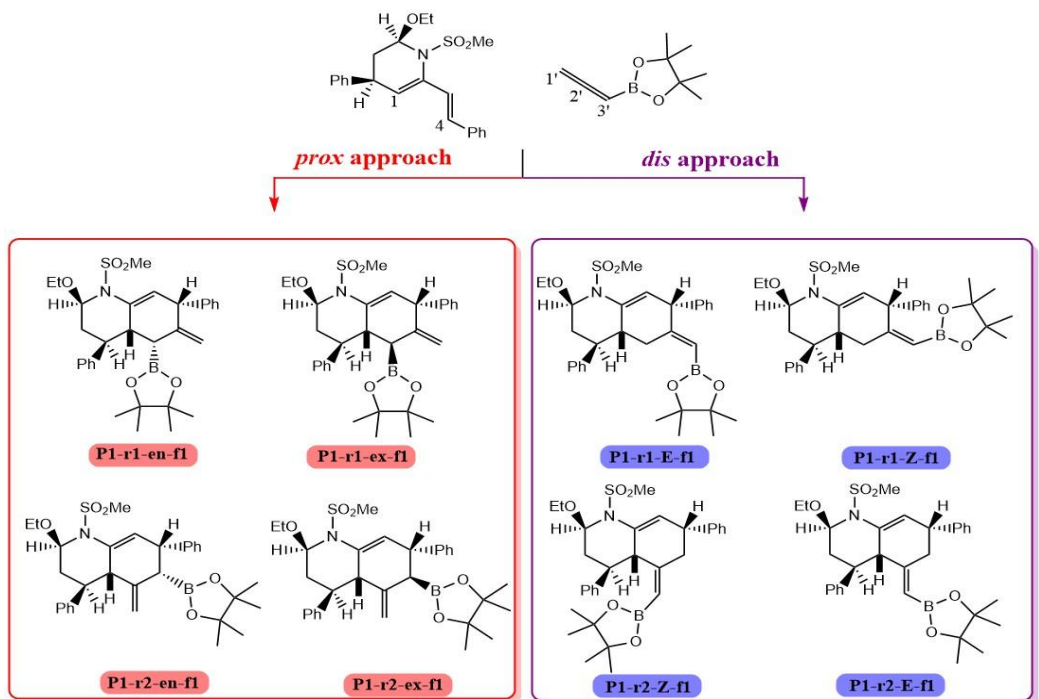

*Figure 2: Potential products studied in the reaction of **VPS** with dienophile **a**. Only possible products obtained from an approach on face f1 are depicted. f2 products can be obtained by switching the stereochemistry at the asymmetric carbons C1 and C4 of VPS.*

The reaction routes have been localized in the gas phase, in toluene and acetonitrile at various temperatures, and the resulting thermodynamic features of the associated products and transition states (TSs) are reported in **Tables S1-15** of the Supporting Information. **Figure 3** shows a diagram of the reaction profile findings in toluene and acetonitrile at 25 °C for the reaction at both the *prox* and *dis* bonds.

In the gas phase, the competing Diels-Alder pathways exhibit a clear separation between kinetic and thermodynamic control. Examination of the computed free energies of activation reveals that the lowest-energy transition state corresponds to a *prox* attack, namely the r2-ex-f1 pathway (TS1-r2-ex-f1, $\Delta G^\ddagger$ = 27.97 kcal.mol$^{-1}$), whereas the lowest-energy product, P1-r1-E-f2 ($\Delta G$ = -38.93 kcal.mol$^{-1}$), arises from a *dis* attack. Thus, under strictly kinetic conditions such as low temperature or rapid quench, the reaction should proceed predominantly via the proximal bond as predicted by the local CDFT indices. We note that the correspondence between the CDFT results and the kinetic product is reasonable since the kinetic conditions depend more on initial interactions of the reactants than on the thermodynamic conditions. Conversely, if allowed to equilibrate, the thermodynamic product will be derived from the distal double bond (P1-r1-E-f2), which is nearly 3 kcal.mol$^{-1}$ more stable than the best *prox* adduct (P1-r1-en-f1, $\Delta G$ = -35.99 kcal.mol$^{-1}$).

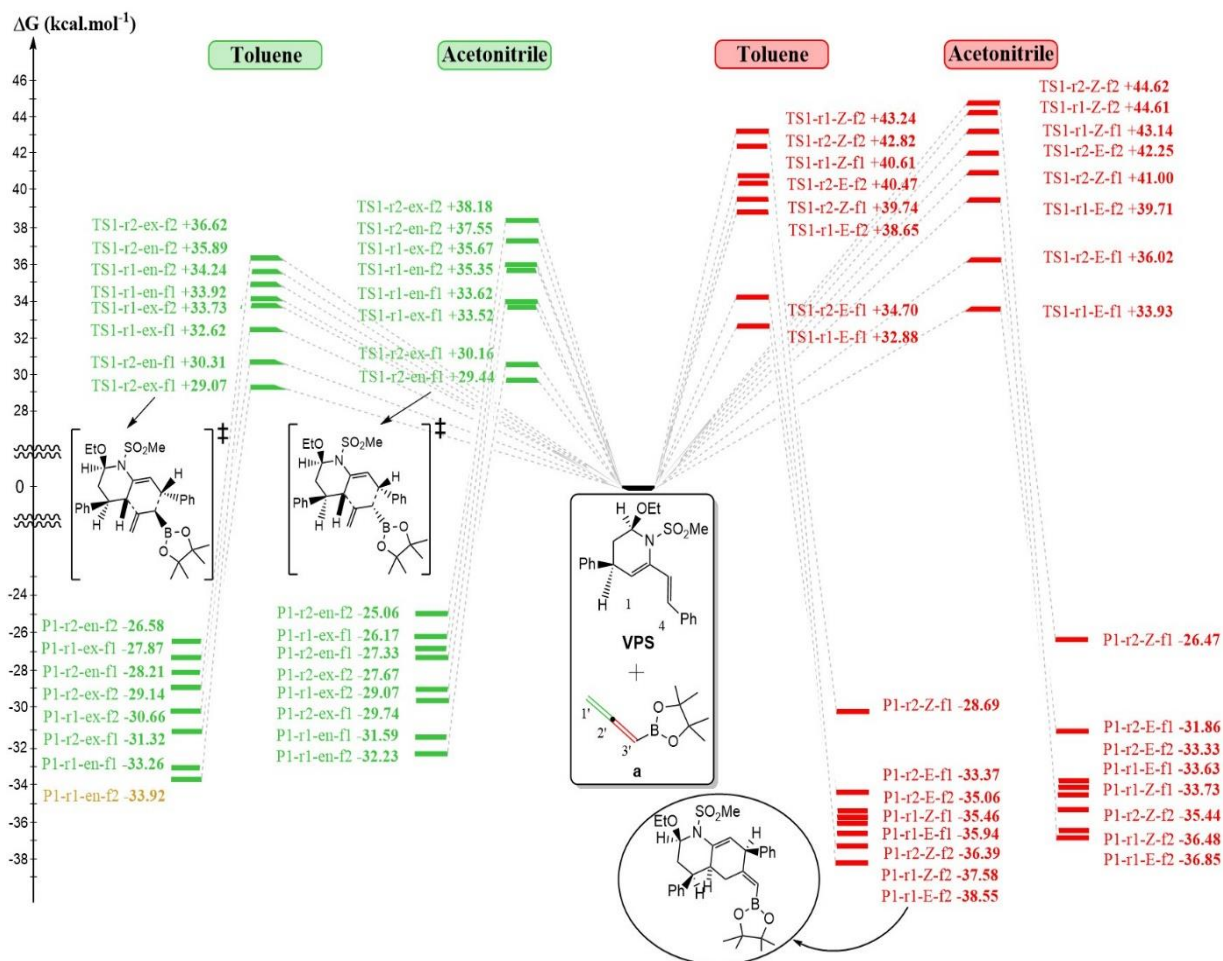

*Figure 3: Reaction profile diagram of the reaction of **VPS** with dienophile **a** at the ωB97X-D/6-311g(d,p) level of theory, in toluene and acetonitrile. Relative energies are given at room temperature in units of kcal.mol$^{-1}$.*

Introduction of the toluene solvent (**Tables S3-S8**) at room temperature subtly attenuates these energetic differences but does not overturn the fundamental preference observed in the gas phase. At 25 °C in toluene, the barrier for TS1-r2-ex-f1 remains lower (ΔG‡ = 29.07 kcal.mol$^{-1}$) than any competing *dis* barrier (lowest is ΔG‡ = 32.88 kcal.mol$^{-1}$ for TS1-r1-E-f1), while the ensemble of *dis* products remains more stable (most stable ΔG = -38.55 kcal.mol$^{-1}$ versus -33.92 kcal.mol$^{-1}$ for *prox* products). In essence, the solvent slightly raises all activation free energies and slightly destabilizes the products, but the kinetic-thermodynamic separation persists with identical periselectivity, regiochemical and stereochemical outcomes. As temperature increases, entropic contributions (-TΔS) progressively diminish the thermodynamic advantage of some of the products. On the one hand, P1-r2-ex-f1 remains the most accessible TS across all temperatures indicating that even at elevated temperature the reaction is likely to remain under kinetic control unless sufficient time is allowed for thermodynamic equilibration. On the other hand, at 70 °C, P1-r2-Z-f2 (ΔG =-36.09 kcal.mol$^{-1}$ ) becomes more stable than P1-r1-E-f2 (ΔG = -34.13 kcal.mol$^{-1}$). However, the superior thermodynamic stability of P4-r1-E-f2 is regained again at 100 °C. The unexpected shift at 70 °C arises because the P1-r2-Z-f2 pathway involves smaller entropy losses upon cycloaddition compared to P1-r1-E-f2 which is destabilized by its larger negative entropy contribution. Thus, while P1-r1-E-f2 is the primary thermodynamic product of the reaction, our findings indicate that under mid-range heating conditions close 70 °C, the P1-r2-Z-f2 will in fact dominate. In acetonitrile (**Tables S9-S14**), the previously observed thermodynamic selectivity towards P4-r1-E-f2 is consistent and there is no switch to the P1-r2-Z-f2 product nor any other one. However, the lowest-barrier pathway changes to TS4-r2-en-f1 instead of the TS1-r2-ex-f1 which dominates in the gas phase and toluene. In general, TS1-r2-ex-f1 is less stable by 0.7-1 kcal.mol$^{-1}$, and the separation in energy between it and TS1-r2-en-f1 increases as temperature rises but at the cost of higher barriers. These barriers evolve from 29.44 kcal.mol$^{-1}$, to 31.85 kcal.mol$^{-1}$, and then 33.47 kcal.mol$^{-1}$ at 25 °C, 70 °C, and 100 °C, respectively.

In summary, our findings indicate that if strict thermodynamic conditions are applied, the reaction will occur at the distal double bond of the dienophile **a** and will be highly selective to P1-r1-E-f2, except near 70 °C in toluene where P1-r2-Z-f2 overtops it in stability. Nevertheless, under the experimentally more convenient kinetic conditions, the reaction will occur at the proximal double bond -in agreement with literature on allenes [30]- and the pathway r2-ex-f1 will dominate the picture in the nonpolar toluene solvent while pathway r2-en-f1 dominates the picture in the polar acetonitrile. Additionally, it is expected to have complete *f1* π-facial selectivity under kinetic control and complete *f2* π-facial selectivity under thermodynamic control. Overall, the findings suggest that there is a high potential for obtaining various isomers provided that a proper optimization of reaction conditions is carried out.

## 2.2. Energy analysis in the reaction of VPS with dienophiles b1 and b2

The reaction of **VPS** with **b1** and **b2** is explored in this section. We first consider the reaction of **VPS** with the chlorocyclohexylvinylborane (**b1**), then we compare the results with the bromine substituted derivative **b2**. We adopt a similar notation for the various stationary points as was done in the reaction of **VPS** and **a**, but omitting the E/Z labels that were used for the geometric configurations. The reaction can proceed according to eight possible pathways; the corresponding products are depicted in **Figure 4**.

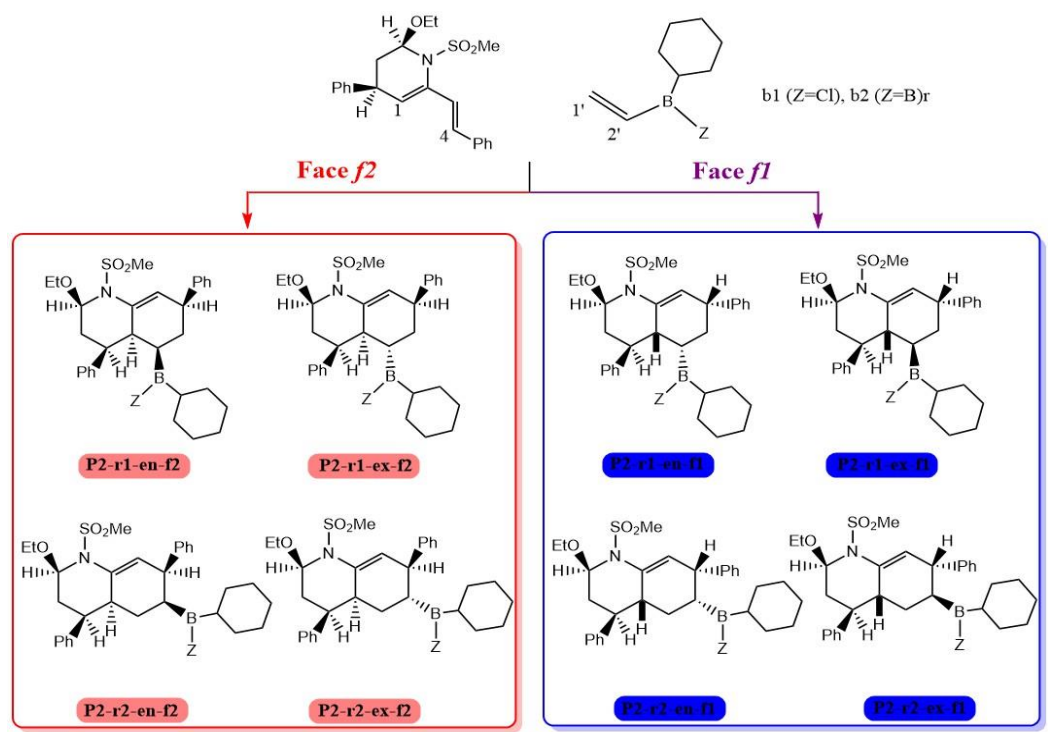

*Figure 4: Potential products studied in the reaction of **VPS** with dienophiles **b1** and **b2**.*

The eight reaction pathways were investigated in the gas phase as well as in toluene and acetonitrile at different temperatures. The corresponding thermodynamic properties of the products and TSs are presented in **Tables S16-S23** and **Tables S24-S30** of the Supporting Information, for **b1** and **b2**, respectively. A diagram summarizing the free activation and reaction energies for **b1** in toluene and acetonitrile at 25 °C is provided in **Figure 5**.

Analysis of the transition state and product free energies decisively points to the *r2* regioisomer. The stereochemical and π-facial outcomes, however, are inverted when comparing the kinetic and thermodynamic pathways.

Across gas-phase and toluene (**Tables S16-S18**), P2-r2-en-f2 emerges as the most stable cycloadduct (ΔG ≈ -28.1 to -22.1 kcal.mol$^{-1}$), always about 0.65 kcal.mol$^{-1}$ lower than its closest competitor which is the corresponding exo product (P2-r2-ex-f2). P2-r1-ex-f2 also closely follows with energy difference of about 0.66 kcal.mol$^{-1}$ compared to P2-r2-en-f2, although it is much less stable in the gas phase. Considering that the gap between P2-r2-en-f2 and its closest competitors remains virtually unchanged with solvent or temperature, an equilibrated mixture will favor the r2-en-f2 isomer even at elevated temperatures. By contrast, the lowest-barrier pathway is via TS2-r2-ex-f1 (ΔG‡ ≈ 22.2 to 26.9 kcal.mol$^{-1}$), roughly 1.4-1.8 kcal.mol$^{-1}$ below the next-most accessible TS (TS2-r2-en-f1). The stability of TS2-r2-ex-f1 in comparison to TS2-r2-en-f1 is enthalpically-driven. This is shown by the bigger differences in relative enthalpy ΔΔH‡ ≈ 5-6 kcal.mol$^{-1}$, and only modest ΔΔS‡ differences. Consequently, it would translate into a strong selectivity for the r2-ex-f1 pathway when the reaction is run irreversibly at low temperature and short duration. Increasing temperature from 25 to 100 °C in toluene systematically makes both products and TSs less stable due to growing -TΔS penalties, but the relative ordering and energy separations remain constant. In general, we expect the reaction to run smoothly at low temperatures. In practice, performing the cycloaddition at low temperature (≤ 25 °C) under irreversible conditions will yield predominantly the kinetic r2-ex-f1 product, whereas allowing the reaction to reach higher temperature or equilibrate it will drive the mixture toward the thermodynamically favored r2-en-f2 isomer, although at a higher energy barrier cost. Concerning selectivity trends in the reaction, it is easily noticeable that both thermodynamic and kinetic preferences overwhelmingly favor *r2* over *r1* and, therefore, we expect exclusive formation of the *r2* regioisomer. Additionally, π-facial preference flips between the TSs (*f1*) and the final, relaxed adducts (*f2*). As for stereoselectivity, it is expected that the exo approach will dominate under kinetic control while, under thermodynamic control, the endo would be favored.

When the chlorine substituent is replaced by bromine, it tunes the gas phase (**Table S22**) stereochemical bias making the r2-ex-f2 adduct slightly more favorable than the r2-en-f2 by -0.36 kcal.mol$^{-1}$ and, therefore, it becomes the thermodynamic product. Nevertheless, the toluene solvent (**Tables S23-S25**) restores the *endo* preference leading again to the r2-en-f2 product observed with the Cl substituent. Other than that, the Br substitution leaves the regio-, stereo- and facial selectivities, as well as the solvent/temperature trends fundamentally intact. The kinetic product always remains P2-r2-ex-f1.

When tested in acetonitrile (**Tables S19-S21** and **Tables S26-S28**), we observe again that across all conditions the same two pathways dominate, r2-ex-f1 under kinetic control, and r2-en-f2 under thermodynamic control. No relevant change is observed in any of the reaction features in comparison to the toluene solvent and therefore both are suitable solvents for this reaction from a polarity perspective.

It is worthy to note that suggested by the initial CDFT analysis, the reaction with **b1** and **b2** shows lower reaction barriers than that with dienophile **a**. Indeed, while all reactions show negative Gibbs free energy along all pathways, the reactions involving **b1** and **b2** are much more stable from a kinetic point of view. Overall, all reactions addressed so far show reaction barriers comparable to known experimental reactions[67, 68].

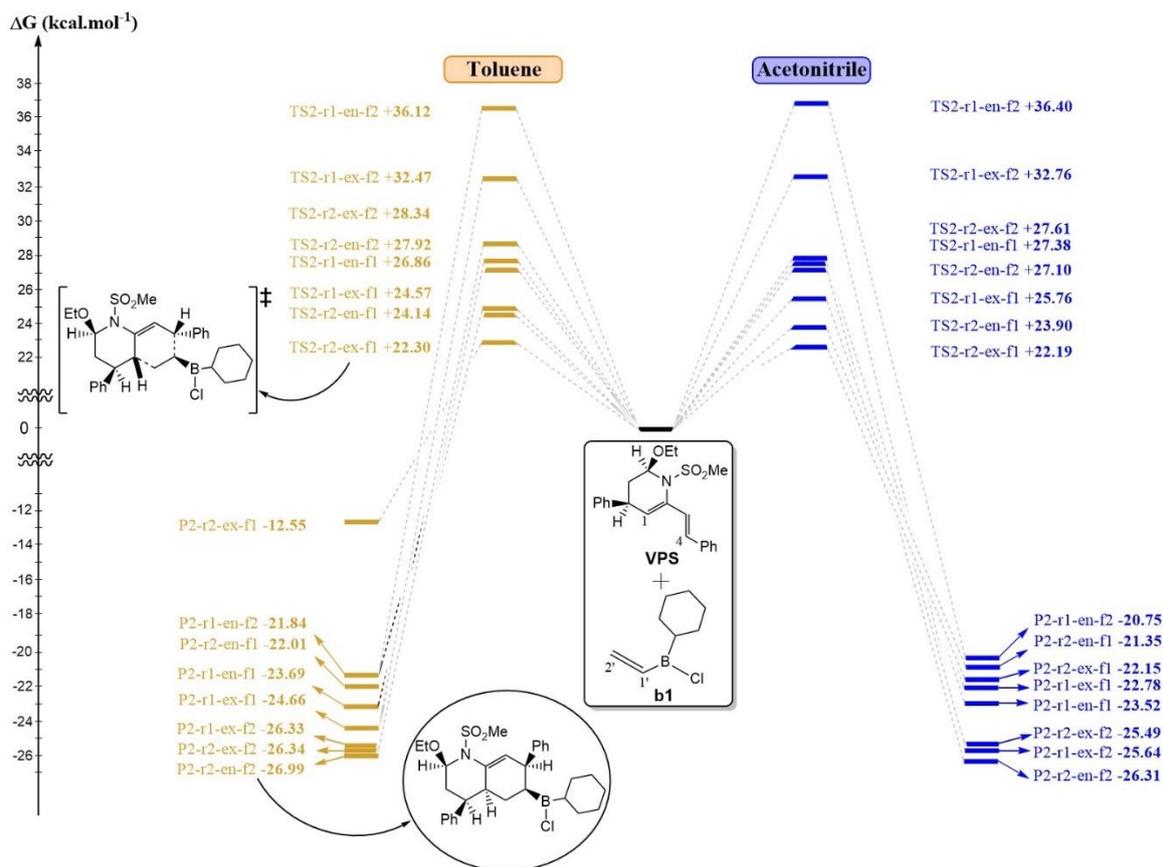

*Figure 5: Reaction profile diagram of the reaction of **VPS** with dienophile **b1** at the ωB97X-D/6-311g(d,p) level of theory, in toluene and acetonitrile. Relative energies are given at room temperature in units of kcal.mol$^{-1}$.*

### 2.3. Energy analysis in the reaction of VPS with dienophile c

The reaction pathways associated with the reaction of **VPS** with dienophile **c** are studied. There are eight possible pathways where the products are shown in **Figure 6**. We adopt the same notation used in the previous reactions for labeling the various stationary points.

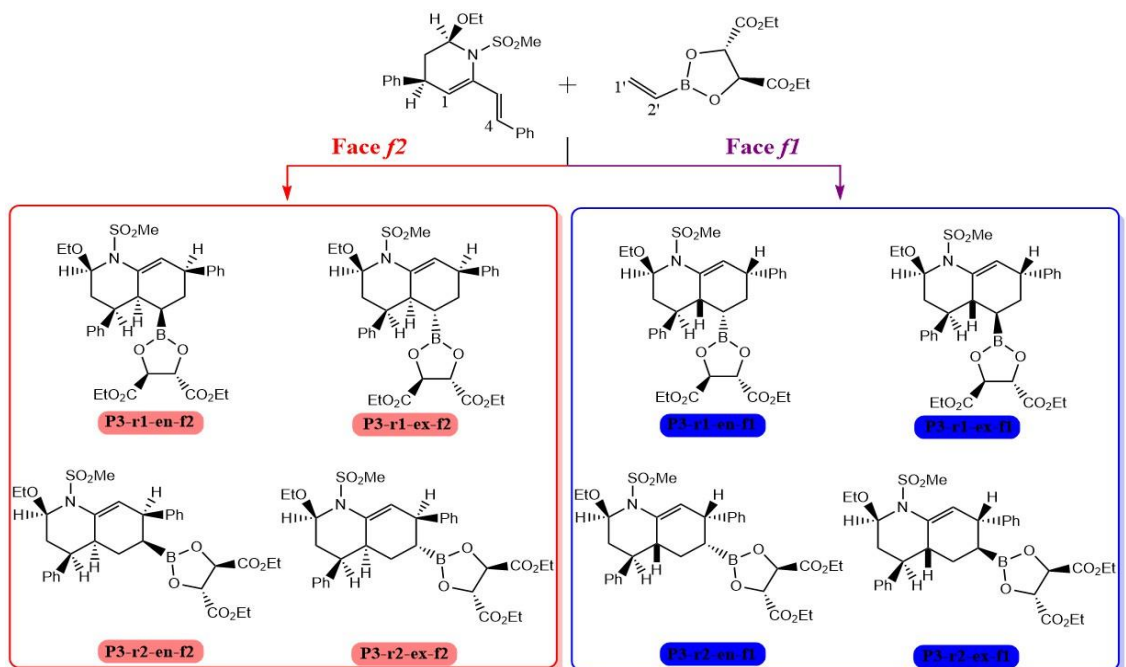

*Figure 6: Potential products studied in the reaction of **VPS** with dienophile **c**.*

The results in the gas phase, toluene and acetonitrile for the possible eight pathways are reported in **Tables S31-S37** of the Supporting Information. The corresponding free energy profiles in toluene and acetonitrile at 25 °C are illustrated in **Figure 7**.

In the gas phase, our eight computed pathways reveal a clear distinction between kinetic and thermodynamic control that will guide our forthcoming solvent- and temperature-dependent analysis. Three pathways, namely r2-ex-f1, r1-en-f1, and r1-ex-f1 show similar activation barriers ($\Delta G^\ddagger \approx$ +25.11, +25.15, and +25.18 kcal.mol$^{-1}$) separated by less than 0.1 kcal.mol$^{-1}$. This predicts that under irreversible conditions, the kinetic product will correspond to one or a mixture of these pathways. By contrast, the global minimum in product free energy is P3-r1-en-f1 ($\Delta G \approx$ -32.83 kcal.mol$^{-1}$), indicating that under thermodynamic control the reaction will ultimately favor the r1-en-f1 isomer. Enthalpic driving force dominates in all cases, with large exothermic $\Delta H$ values offset by moderate entropic penalties (T$\Delta S \approx$ +5-6 kcal.mol$^{-1}$ at 25 °C), so $\Delta G$ trends closely mirror the $\Delta H$ ones. Across every regio- and stereochemical variant, attack on *f1* is more favorable than *f2* at the level of the TSs, suggesting reduced steric strain and/or potential secondary interactions on that π-face. Although, the strain seems to relax in most cases (except for r1-en-f1) so that the *f2* products are more stable than the *f1* ones. These benchmarks in the absence of solvent set the stage for exploring how toluene, acetonitrile, and varying temperatures will modulate both barrier heights and product stabilities, potentially altering the balance between the kinetic and thermodynamic pathways.

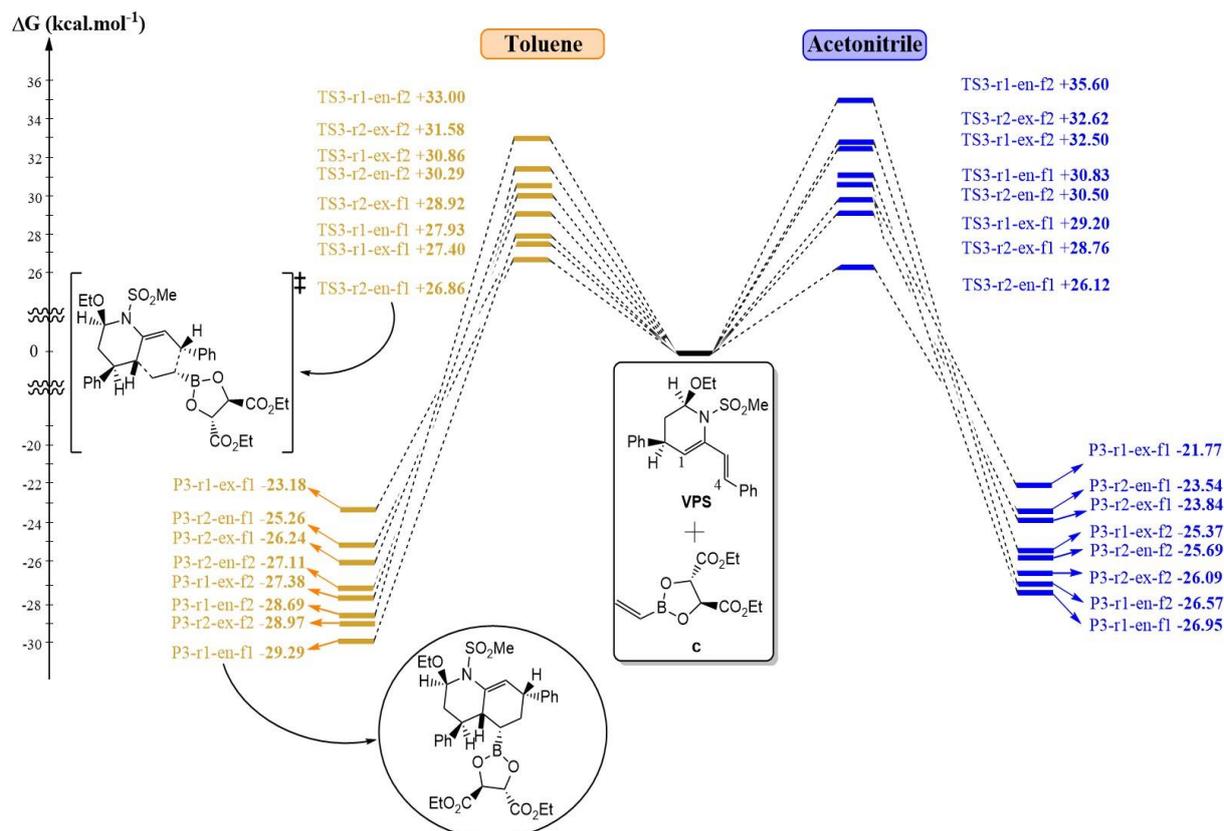

*Figure 7: Reaction profile diagram of the reaction of **VPS** with dienophile **c** at the ωB97X-D/6-311g(d,p) level of theory, in toluene and acetonitrile. Relative free energies are given at room temperature in units of kcal.mol$^{-1}$.*

In toluene (**Tables S32-S32**) at 25 °C, the overall landscape of regio-, stereo-, and π-facial selectivity remains qualitatively similar to the gas phase with respect to the thermodynamic product, but with modest energetic leveling due to solvent stabilization. The r1-en-f1 isomer remains the thermodynamic minimum (ΔG = -29.29 kcal.mol$^{-1}$), albeit ~3.5 kcal.mol$^{-1}$ less exergonic than in vacuo, and closely followed by the P3-r2-ex-f2 isomer which is less stable by 0.32 kcal.mol$^{-1}$. π-Facial preference for attack on face *f1* persists across all pathways as shown by the lower activation barriers, and enthalpy still dominates the free-energy trends (large exothermic offset by ~+5.5 kcal.mol$^{-1}$ of -TΔS at 25 °C ). However, the kinetic picture shifts in toluene where the lowest ΔG‡ (26.86 kcal.mol$^{-1}$) is for the TS3-r2-en-f1 whereas the gas-phase barriers indicate a mixture of products (r2-ex-f1, r1-en-f1, and r1-ex-f1) and there is no decisive selectivity. Thus, under kinetic conditions, toluene drives the system to the r2-en-f1 pathway. Considering the energetic gaps obtained thus far, we expect that increasing temperatures can alter the energetic order of the stationary points. As the temperature in toluene rises from 25 °C to 70 °C and then to 100 °C, the Diels-Alder landscape remains mostly unchanged from a kinetic perspective, but thermodynamic balance undergoes an important change. Indeed, P3-r2-ex-f2, the second-most stable product, gains in stability at 70 °C and 100 °C. In particular, it approaches the thermodynamic minimum P3-r1-en-f1 by about 0.1 kcal.mol$^{-1}$ at 100 °C which implies that at high temperature we should expect a mixture of the two products. Overall, it is predicted that, under typical conditions in toluene with ample reaction time to promote reversibility, the mixture will equilibrate to the thermodynamic product, yielding P3-r1-en-f1 as the major product. Furthermore, if the temperature is raised toward 100 °C, P3-r2-ex-f2 should also appear as a minor product. Alternatively, if the reaction is run at low temperatures, and conditions that allow back-reaction are avoided, the kinetic preference for the lower-barrier pathway will dominate. In that scenario, P3-r2-en-f1 forms the fastest and will be the primary adduct.

In acetonitrile (**Tables S35-S37**), the thermodynamic product observed in the gas phase and in toluene, i.e., P3-r1-en-f1, is once again favored. In addition, the second-most stable thermodynamic product is now P3-r1-en-f2 replacing P3-r2-ex-f2. On the kinetic level, similar to toluene, P3-r2-en-f1 is the most stable. However, a notable effect of acetonitrile is that the energy gap between TS3-r2-en-f1 and the second most stable TS is significantly increased to about 2.6 kcal.mol$^{-1}$ for all temperature suggesting complete selectivity, towards P3-r2-en-f1. Consequently, acetonitrile is an excellent choice to obtain the kinetic product.

## 3. Global Electron Density Transfer Analysis

GEDT is a valuable theoretical descriptor used to quantify how much electron density shifts from a nucleophilic to an electrophilic partner during the formation of a transition state in a chemical reaction. Forward Electron Density Flux (FEDF) and Reverse Electron Density Flux (REDF) are two complementary ways to classify polar organic reactions by the direction of electron-density flow at the transition state, as measured by the GEDT. In FEDF processes, the diene acts as the nucleophile, donating electron density into the dienophile, so the GEDT is positive from the diene toward the partner. In contrast, REDF reactions feature the dienophile as the nucleophile and the diene as the electrophile, reversing the usual electron-flow direction so that the GEDT is positive from the smaller π-system toward the larger one. Table 3 contains the calculated GEDT for all main reactions.

In the reaction of **VPS** and **a**, all TSs underscore an FEDF scenario where electron density flows from the diene nucleophile toward the dienophile electrophile. The TSs show distinct GEDT values ranging from near zero transfer to 0.21 e. For the reaction at the proximal double bond of **a**, the highest transfer occurs at TS1-r1-en-f2 (0.15 e), closely followed by TS1-r1-ex-f2 (0.13 e), while for the reaction at the distal double bond of **a**, TS1-r2-Z-f2, shows the largest GEDT at 0.21 e, with TS1-r2-E-f2 close behind at 0.18 e. Some pathways like r1-en-f1 (0.0032 e), r1-ex-f1 (0.0066 e), r1-E-f2 (0.0016 e) or r2-Z-f1 (0.01 e) exhibit virtually no density transfer and are practically non-polar reactions.

*Table 3: Global Electron Density Transfer for the reaction of **VPS** with dienophiles **a**, **b1**, **b2**, and **c**. All values are in units of e. The label KP indicates the kinetic product according to the calculated activation barriers.*

| VPS + a (Proximal bond) | | VPS + a (Distal bond) | |
|---|---|---|---|
| **TS1-r1-en-f1** | 0.003 | **TS1-r1-E-f1** | 0.04 |
| **TS1-r1-en-f2** | 0.15 | **TS1-r1-E-f2** | 0.002 |
| **TS1-r1-ex-f1** | 0.01 | **TS1-r1-Z-f1** | 0.03 |
| **TS1-r1-ex-f2** | 0.13 | **TS1-r1-Z-f2** | 0.02 |
| **TS1-r2-en-f1** | 0.07 | **TS1-r2-E-f1** | 0.02 |
| **TS1-r2-en-f2** | 0.08 | **TS1-r2-E-f2** | 0.18 |
| **TS1-r2-ex-f1** | 0.08 | **TS1-r2-Z-f1** | 0.01 |
| **TS1-r2-ex-f2** | 0.04 | **TS1-r2-Z-f2** | 0.21 |
| VPS + b1/b2 | | VPS + c | |
| **TS2-r1-en-f1** | 0.06/0.06 | **TS3-r1-en-f1** | 0.01 |
| **TS2-r1-ex-f1** | 0.06/0.06 | **TS3-r1-ex-f1** | 0.01 |
| **TS2-r2-en-f1** | 0.15/0.15 | **TS3-r2-en-f1** | 0.02 |
| **TS2-r2-ex-f1** | 0.11 /0.11 | **TS3-r2-ex-f1** | 0.03 |
| **TS2-r1-en-f2** | 0.13/0.14 | **TS3-r1-en-f2** | 0.06 |
| **TS2-r1-ex-f2** | 0.11/0.11 | **TS3-r1-ex-f2** | 0.04 |
| **TS2-r2-en-f2** | 0.18/0.19 | **TS3-r2-en-f2** | 0.07 |
| **TS2-r2-ex-f2** | 0.15 /0.16 | **TS3-r2-ex-f2** | 0.08 |

In the reaction of **VPS** with **b1** and **b2**, GEDT values for b1/b2 at the eight TSs span from 0.06/0.06 e up to 0.18/0.19 e, confirming a forward electron-density flux (FEDF) from the diene toward the dienophile. The kinetic product, TS2-r2-ex-f1, shows a moderate GEDT of 0.11 e, and surprisingly the largest transfer (0.18/0.19 e) is observed for TS2-r2-en-f2 which is the thermodynamic product. At 0.18/0.19 e, the pathway can be roughly considered a polar route, yet the associated barrier is in fact higher than most of the lower GEDT pathways. This is interesting considering that higher GEDTs are usually correlated with lower barriers in polar processes [69]. Nonetheless, non-polar and low-polarity DA reactions still remain reliable synthetic routes, especially with proper solvation and heating conditions [70, 71].

Conversely, in the reaction of **VPS** and **c**, overall GEDT values are small (0.01-0.08 e), reflecting a low polarity Diels-Alder process. The kinetic product, TS3-r2-en-f1, has a GEDT of 0.02 e, higher than the tiny transfer to the thermodynamic product, TS3-r1-en-f1 (0.01 e). The pathway with the highest GEDT (0.08 e) corresponds to TS3-r2-ex-f2.

Overall, given the discrepancy with the result from the calculated energies for the various TSs, we expect that apart from electron-density flux derived from the natural charges, other effects such as steric hindrance play an important role in the TSs stability in the various pathways. This prompts us to investigate the underlying steric effects taking place at the level of the TSs.

## Conclusions

In this work we have introduced a distinct approach to access boron-functionalized octahydroquinolines via a diene-transmissive hetero-Diels-Alder protocol. It is computationally demonstrated that the procedure can be carried out with a pyridine-based diene with a range of possible substitutions and three possible boron-containing dienophiles, namely, allenylboronic acid pinacol ester, alkylhalovinylboranes, and 4S,5S-[Bis(carbethoxy)]-2-ethenyl-1,3,2-dioxaborolane. All studied reactions exhibit negative Gibbs free energies, indicating thermodynamically favorable processes, and feature reasonable energy barriers comparable to those of known experimentally viable Diels–Alder reactions. Among the dienophiles, alkylhalovinylboranes are predicted to be the most efficient. In general, a variety of boronated products are accessible and the peri-, regio-, stereo- and diastereo-selectivity were all elucidated and some insights on how to control them with solvation and heat are provided. The determined reaction pathways as well as topological studies show that the reactions mechanisms, in the kinetically most stable pathways, is the standard asynchronous one-step mechanism. To our knowledge, this connection between a diene-transmissive hetero-Diels-Alder reaction and boronated products has not been investigated before. Our results have shown this strategy holds significant potential as an approach to obtain boron-derivatives of the octahydroquinoline family. The insights on selectivity and mechanism are expected to accelerate experimental efforts and significantly expand the chemical space available to synthetic and medicinal chemists alike.

## Acknowledgements

This work has received funding from the European Union's Horizon 2020 research and innovation program under Marie Sklodowska-Curie grant agreement No. 872081, and grant PID2022-136228NB-C21 funded by MICIU/AEI/10.13039/501100011033 and, as appropriate, by ERDF A way of making Europe, by ERDF/EU, by the European Union or by the European Union NextGenerationEU/PRTR. This work is also supported by the Consejería de Transformación Económica, Industria, Conocimiento y Universidades, Junta de Andalucía and European Regional Development Fund (ERDF 2021-2027) under the project EPIT1462023. The authors acknowledge the Moroccan Association of Theoretical Chemists for providing the computational programs.

## Associated Content

Results not included in the manuscript are available in the Supporting Information. These include calculations at different temperatures, additional IGM results, and Infrared spectra of the most stable products. Cartesian coordinates (in Å) of the most stable products are also provided.